# Mass-Resolved Electronic Circular Dichroism Ion Spectroscopy

Steven Daly,[1] Frédéric Rosu,[2] Valérie Gabelica[1]*

[1]Université de Bordeaux, Inserm & CNRS, Laboratoire Acides Nucléiques: Régulations Naturelle et Artificielle (ARNA, U1212, UMR5320), IECB, 2 rue Robert Escarpit, 33607 Pessac, France.

[2] Université de Bordeaux, CNRS & Inserm, Institut Européen de Chimie et Biologie (IECB, UMS3033, US001), 2 rue Robert Escarpit, 33607 Pessac, France.

*Correspondence to: v.gabelica@iecb.u-bordeaux.fr.

**Abstract:** DNA and proteins are chiral: their three-dimensional structure cannot be superimposed with its mirror image. Circular dichroism spectroscopy is widely used to characterize chiral compounds, but data interpretation is difficult in the case of mixtures. We recorded for the first time the electronic circular dichroism spectra of DNA helices separated in a mass spectrometer. We electrosprayed guanine-rich strands having various secondary structures as negative ions, irradiated them with a laser, and measured the difference in electron photodetachment efficiency between left and right circularly polarized light. The reconstructed circular dichroism ion spectra resemble the solution ones, thereby allowing us to assign the DNA helical topology. The ability to measure circular dichroism directly on biomolecular ions expands the capabilities of mass spectrometry for structural analysis.

Two centuries ago, the interaction of polarized light with crystals revealed that many molecules come in two forms, non-superimposable with their mirror image (*1*). The activity and toxicity of natural or synthetic molecules often depend on their chirality (*2*). Identifying which mirror image is present in a sample is thus crucial. Furthermore, because biomolecules such as DNA or proteins consist of repetitive chiral subunits, they can form helical structures, as in the iconic DNA double helix (*3*). Characterizing the types of helical structures formed by biomolecules is therefore also essential for structural biology.

Mass spectrometry (MS) is a widely used analytical method, with expanding impact in structural biology (*4, 5*). MS excels at separating and quantifying complex mixtures, and additional structural characterization can be obtained using tandem mass spectrometry (*6*), ion mobility spectrometry (*7*) or infrared ion spectroscopy (*8-10*). However, mass spectrometry-based measurements are typically blind to chirality. Characterizing chiral compounds by mass spectrometry currently requires a physical interaction with other chiral molecules (*11*), either by separation on chiral phases in front of the mass spectrometer or by forming complexes with chiral auxiliaries in the gas phase (*12*).

Another way to characterize chiral biomolecules is through interaction with chiral light. In solution, DNA and proteins are conveniently characterized by circular dichroism (CD), which measures the difference of absorption between left- and right-circularly polarized light. Electronic CD is particularly useful to characterize the different types of helices formed by nucleic acids (*13*) or the secondary structures formed by proteins (*14*). The structural interpretation relies on spectral databases of known structures. However, the interpretation of CD spectra becomes difficult for samples wherein multiple species or structures coexist, because the resulting spectrum is the weighted average of all contributions.

Here we report how to record electronic circular dichroism spectra on DNA helices separated in a mass spectrometer. Although CD and MS were combined before (*15-17*), the previous methods were based on resonant multiphoton ionization (REMPI), and were thus applicable only to volatile neutral small molecules (< 200 Da). Here we used electrospray to produce gas phase polyanions of intact DNA multi-helices (> 1000 Da), and interrogated them with chiral UV light directly inside the mass spectrometer.

Our mass analyzer is a quadrupole ion trap (Paul trap) with two opposite holes (1.7 mm diameter) in the ring electrode. We reasoned that this configuration would minimize risks of reflection of the laser beam inside the mass spectrometer, which could possibly alter the circular polarization. The electrospray source was operated in the negative mode to produce multiply charged anions. We used gentle ion transfer conditions to prevent gas-phase restructuring and thus maximize the chances of preserving the solution secondary structures.

We analyzed several DNA G-quadruplex tetra-helical structures (parallel right-handed tetramer TG4T [(dTGGGGT)$_4$•(NH$_4^+$)$_3$], parallel left-handed intramolecular ZG4 [dT(GGT)$_4$TG(TGG)$_3$TGTT•(NH$_4^+$)$_3$] (*18*), right-handed antiparallel intramolecular 5YEY [d(GGGTTA)$_2$GGGTTTGGG•(K$^+$)$_2$] (*19*)) and the silver-mediated parallel G-duplex GAgG ([(dG$_{11}$)$_2$•(Ag$^+$)$_{11}$] (*20*)). These structures were chosen because they have remarkably different electronic CD spectra in solution in the range 220-300 nm (Fig. S1). Also, these structures are stabilized by spines of central cations, which maximizes the likelihood that the G-rich secondary structures will be preserved in the gas phase. The preservation of the known solution shapes was confirmed by the good agreement between helium momentum transfer collision integrals measured in ion mobility spectrometry and those calculated for gas-phase optimized structures (Fig. S2). For TG4T, infrared ion spectroscopy also supported the preservation of G-quartets (*21*). We can thus safely assume that these molecular systems have the same base stacking arrangement in the gas phase as in solution.

When coupled to mass spectrometry, ion spectroscopy is an action spectroscopy: one records the effect of the laser irradiation on the ions. Another reason for selecting DNA as the first test case is that, when irradiating DNA polyanions between 220-300 nm, the resulting action is electron photodetachment (ePD) (*22, 23*). This action has two advantages compared to fragmentation: (i) there are few product ions to quantify, which in terms of statistics will increase the chances to detect significant differences in product ion yields between the two polarizations, and (ii) ePD is monophotonic (*24*), and thus the normalization for fluctuations of the laser power is simply linear for all wavelengths. We verified this linear relationship, and for all spectroscopy experiments we selected the pulse energy ranges in which only linear electron photodetachment was observed (Fig. S3).

Nanosecond laser pulses were generated using a GWU Premiscan OPO pumped by a Spectra Physics PRO-230-30 Nd:YAG. To generate circularly polarized laser pulses, the beam passes through an air-spaced Rochon prism (Kogakugiken, Japan), which gives pure linearly polarized light. Next, the laser passes through an achromatic broadband quarter wave plate (Kogakugiken, Japan) (Fig. 1, more details in supplementary materials and methods and Figs. S4–S6). The angle of the fast axis of the quarter wave plate relative to the Rochon prism determines the final polarization state of the laser pulse. The laser is coupled to the ion trap via a fused silica window in the vacuum manifold. The percentage of circular polarization was greater than 95% (Fig. S7, Table S1). The ions are randomly oriented inside the Paul trap.

To measure their circular dichroism, the ions are mass-selected, irradiated by a single laser pulse of the selected wavelength, polarization and pulse energy, then mass analyzed (Fig. 1). Mass spectra and pulse energy are averaged for 90 seconds. The polarization state of the laser pulse is then changed by rotating the quarter wave plate, and the mass spectrum and pulse energy acquired for another 90 seconds. This process is repeated 10 times for each polarization. The relative electron detachment yield for each mass spectrum is calculated as follows:

$$Y_{ePD} = \left(\frac{I_{ePD}}{I_{total}}\right)/\lambda E_{trans} \qquad (1)$$

$I_{ePD}$ is the integrated intensity of the peak due to electron detachment, $I_{total}$ is the total intensity, $\lambda$ is the wavelength, and $E_{trans}$ is the transmitted pulse energy determined from the reflected pulse energy by calibration (Fig. S8). We then calculate the average value of the relative electron detachment yield for left- and right-handed circular polarizations. The circular dichroism monitored by electron photodetachment is expressed as an asymmetry factor, approximated by:

$$g_{ePD} = \frac{\Delta ePD}{ePD} = 2 * \left(\frac{\overline{Y_{ePD}^{LCP}} - \overline{Y_{ePD}^{LCP}}}{\overline{Y_{ePD}^{LCP}} + \overline{Y_{ePD}^{LCP}}}\right) \qquad (2)$$

The gas-phase CD spectra are reconstructed by plotting the asymmetry factor $g_{ePD}$ as a function of the wavelength. To facilitate the visual comparison, the solution CD spectra are plotted as $\Delta A/\bar{A}$ (*A* being the absorbance).

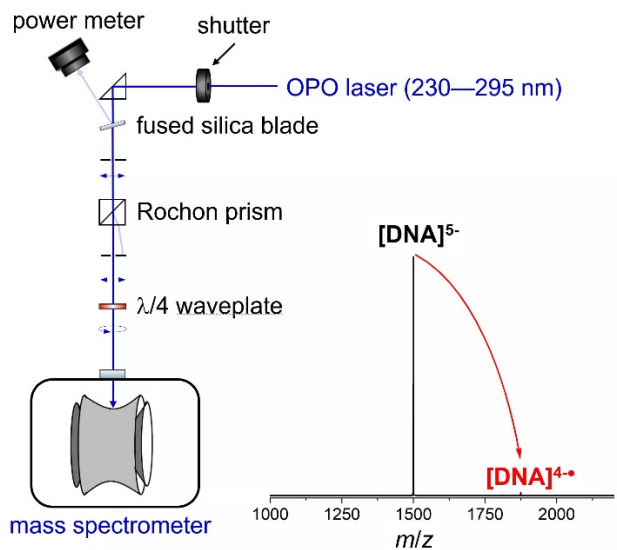

**Fig. 1.** Experimental setup for generating circularly polarized laser pulses (full details in supporting information), and typical mass spectrum following isolation of D-[(dTGGGGT)$_4$.(NH$_4^+$)$_3$]$^{5-}$ ions and irradiation with left circularly polarized (LCP) light at 260 nm.

To unambiguously prove that the gas-phase CD effect comes from the sample and not from an instrumental artifact, we performed the experiment on D-[(TGGGGT)$_4$•(NH$_4^+$)$_3$] formed from the natural DNA backbone (all D-sugars), and with its enantiomer (all L-sugars). The gas-phase CD signals have opposite signs (Fig. 2A). The magnitude is not exactly reversed, however. At 260 nm, we recorded a series of CD measurements by varying the relative concentration of the two enantiomers (D:L) in solution from only D to only L (Fig. 2A, inset). We obtained a straight line (r²=0.975), showing that the differences in ePD with left- and right-circularly polarized light are reporting on circular dichroism and suggesting that the technique could be suitable for quantification. The line did not pass exactly through 0 when the proportions were 50:50, indicating the presence of residual instrumental artifacts, likely due the fact that the polarization is only ~95% circular. In the future, using an achiral internal standard could further reduce the uncertainty. Furthermore, if such imperfections vary with the wavelength, it would induce some distortion in the CD spectra that might hamper structural assignment. We thus compared the solution and gas-phase CD spectra for other typical DNA helices (Fig. 2B—D; symbols for gas-phase CD, lines for solution CD). The solution and gas phase CD spectral shapes are very similar. This suggests that the following conditions are all met: (i) The base stacking pattern existing in solution is preserved in the gas-phase ions. (ii) The gas-phase action reflects the absorption, in terms of sensitivity to the circular polarization. In other words, the electronic excited states that are responsible for the CD effect also trigger electron photodetachment. (iii) Possible distortions due to imperfections in the polarization do not preclude assigning the base stacking arrangement based on the spectral shape. As a result, the gas-phase CD spectra can unambiguously discriminate between the different guanine-rich oligonucleotide structures.

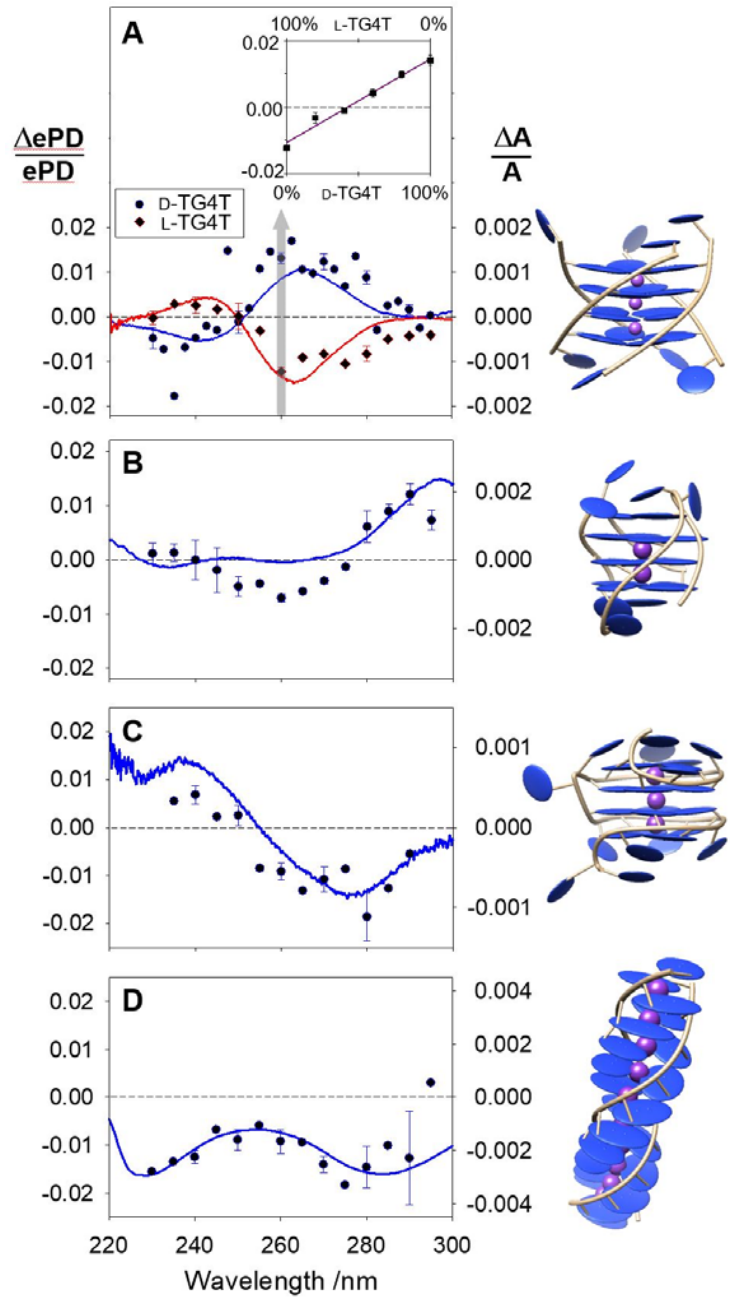

**Fig. 2.** Gas-phase circular dichroism spectra (ΔePD/ePD, symbols) compared to solution phase ones (ΔA/A lines). (A) TG4T: natural right-handed D-[(TGGGGT)$_4$•(NH$_4^+$)$_3$]$^{5-}$ (blue) and its mirror image L- [(TGGGGT)$_4$•(NH$_4^+$)$_3$]$^{5-}$ (red). Inset: Gas phase circular dichroism measured at 260 nm for [(TGGGGT)$_4$•(NH$_4^+$)$_3$]$^{5-}$ ions prepared from 5 µM solutions with varying ratios of D-TG4T and L-TG4T. Error bars show the 95% confidence interval, calculated from the standard error (3 replicas, except at 260 nm with 5 replicas). (B—D) gas-phase and solution phase CD spectra for the antiparallel G-quadruplex 5YEY$^{5-}$ (B), the left-handed G-quadruplex ZG4$^{6-}$ (C) and the G-duplex GAgG$^{5-}$ (D). The illustrations showing helicity were generated using Chimera from coordinated deposited in the protein data back (entries 2O4F, 5YEY, 6GZ6, and 4JRD, respectively), and the gas-phase optimized structures are shown in Fig. S2.

The gas and solution phase asymmetry factors mainly differ by their magnitude, which is consistently larger in the gas phase. This has been observed previously for small neutral molecules measured by REMPI (*16, 25-27*), where magnitudes of up to 25% have been observed (*27*). High CD signals are also observed in photoelectron circular dichroism (PECD) (*28, 29*), but in that case the measured property is the angular distribution of the photoelectrons. Here, we measure the electron detachment yield, and ePD results from a resonant excitation above the detachment threshold (*23, 24*). Given that solution-phase spectra result from the CD in absorption, we conclude that the gas-phase CD effect in photodetachment yields is also due to the resonant absorption of chiral light by a chiral molecule. It is thus puzzling that the CD spectral shapes are similar although the magnitudes differ.

Let us discuss some possible explanations. In small molecules, the solvent can affect the magnitude of circular dichroism (*30*), and thus changes in the dielectric environment from water to vacuum may change the magnitude of the CD spectrum. To test this hypothesis, we recorded solution CD spectra of D-TG4T in mixtures of water up to 50% isopropanol (dielectric constant varying from 80.4 to 44.3). The mass spectra show that the G-quadruplex is intact, and no higher-order structure is present (Fig. S9). No significant change in the value of $\Delta A/\bar{A}$ was observed (Fig. S10), suggesting the dielectric environment plays only a small role. We also recorded the gas-phase CD for the 5-, 6- and 7- ions of D-TG4T, reasoning that if intramolecular electric dipoles influence the electron ejection dynamics, the effect would increase with the charge state. We observe no significant change in either magnitude or shape of the CD spectrum as a function of the charge state (Fig. S11). We thus hypothesize that the origin of the larger gas-phase magnitude comes from the different definition of asymmetry factors. In the gas phase, absorption is revealed by electron detachment. If not all states that absorb result in electron detachment (*24, 31*) while most states that are responsible for CD do, the denominator of the asymmetry factor is smaller in gas phase CD action spectroscopy, and thus the asymmetry factor is larger. The electronic states responsible for the CD effect being most likely to be delocalized on the entire DNA helices, this may result in more efficient auto-detachment after resonant excitation. Thus, although future work is needed to elucidate the origins and dynamics of electron photodetachment from DNA polyanions, for example by measuring the photoelectron circular dichroism, the electron photodetachment action serendipitously revealed itself especially well-suited to probe DNA higher-order structures by ion spectroscopy.

In conclusion, we demonstrated that it is possible to measure the circular dichroism of large DNA polyanions in the gas phase, after separation in a mass spectrometer. The similarity between solution and gas-phase spectra allowed us to distinguish DNA secondary structures, thereby extending the scope and capabilities of structural mass spectrometry. This first demonstration of feasibility opens many avenues towards the study of protein secondary structure with electron photodetachment, or even more diverse classes of chiral molecules with cold mass tagging, while leveraging the separation capabilities of contemporary mass spectrometry.

**References:**

1. F. Arago, *Mém. Class Sci. Math. Phys. Inst. Impérial France* **1**, 93-164 (1811).
2. S. W. Smith, *Toxicol. Sci.* **110**, 4-30 (2009).
3. J. D. Watson, F. H. C. Crick, *Nature*, 737 (1953).
4. I. Liko, T. M. Allison, J. T. Hopper, C. V. Robinson, *Curr. Opin. Struct. Biol.* **40**, 136-144 (2016).
5. P. Lossl, M. van de Waterbeemd, A. J. Heck, *EMBO J.* **35**, 2634-2657 (2016).
6. F. W. McLafferty, *Science* **214**, 280-287 (1981).
7. D. E. Clemmer, M. F. Jarrold, *J. Mass Spectrom.* **32**, 577-592 (1997).
8. E. Garand *et al.*, *Science* **335**, 694-698 (2012).
9. J. Seo *et al.*, *Nat. Chem.* **9**, 39-44 (2017).
10. M. Z. Kamrath, T. R. Rizzo, *Acc. Chem. Res.* **51**, 1487-1495 (2018).
11. H. Awad, A. El-Aneed, *Mass Spectrom. Rev.* **32**, 466-483 (2013).
12. W. A. Tao, F. C. Gozzo, R. G. Cooks, *Anal. Chem.* **73**, 1692-1698 (2001).
13. J. Kypr, I. Kejnovska, D. Renciuk, M. Vorlickova, *Nucleic Acids Res.* **37**, 1713-1725 (2009).
14. A. J. Miles, B. A. Wallace, *Chem. Soc. Rev.* **35**, 39-51 (2006).
15. R. Li, R. Sullivan, W. Al-Basheer, R. M. Pagni, R. N. Compton, *J. Chem. Phys.* **125**, 144304 (2006).
16. U. Boesl von Grafenstein, A. Bornschlegl, *ChemPhysChem* **7**, 2085-2087 (2006).
17. A. Hong *et al.*, *Angew. Chem. Int. Ed.* **53**, 7805-7808 (2014).
18. W. J. Chung *et al.*, *Proc. Natl. Acad. Sci. USA* **112**, 2729-2733 (2015).
19. C. Liu *et al.*, *Chem. Sci.* **10**, 218-226 (2018).
20. S. M. Swasey, F. Rosu, S. M. Copp, V. Gabelica, E. G. Gwinn, *J. Phys. Chem. Lett.* **9**, 6605-6610 (2018).
21. V. Gabelica *et al.*, *J. Am. Chem. Soc.* **130**, 1810-1811 (2008).
22. V. Gabelica *et al.*, *Anal. Chem.* **78**, 6564-6572 (2006).
23. F. Rosu *et al.*, *J. Phys. Chem. A* **116**, 5383-5391 (2012).
24. S. Daly, M. Porrini, F. Rosu, V. Gabelica, *Faraday Discuss.* **217**, 361-382 (2019).
25. P. Horsch, G. Urbasch, K. M. Weitzel, D. Kroner, *Phys. Chem. Chem. Phys.* **13**, 2378-2386 (2011).
26. H. G. Breunig *et al.*, *ChemPhysChem* **10**, 1199-1202 (2009).
27. J. Lepelmeier, K. Titze, A. Kartouzian, U. Boesl, U. Heiz, *ChemPhysChem* **17**, 4052-4058 (2016).
28. L. Nahon, G. A. Garcia, C. J. Harding, E. Mikajlo, I. Powis, *J. Chem. Phys.* **125**, 114309 (2006).
29. M. H. Janssen, I. Powis, *Phys. Chem. Chem. Phys.* **16**, 856-871 (2014).
30. A. R. de Souza, V. F. Ximenes, N. H. Morgon, in *Stereochemistry and Global Connectivity: The Legacy of Ernest L. Eliel Volume 2*. (2017), pp. 91-101.
31. V. Gabelica *et al.*, *J. Am. Chem. Soc.* **129**, 4706-4713 (2007).

**Acknowledgments:** We thank Nina Khristenko for contributing to preliminary experiments. **Funding:** European Research Council, ERC-2013-CoG-616551-DNAFOLDIMS; **Author contributions:** FR and VG conceived the project; VG acquired the funding; FR and SD developed the methodology; SD and FR conducted the research and analyzed the data; SD and VG wrote the paper.

Supplementary Information for:

# Mass-Resolved Electronic Circular Dichroism Ion Spectroscopy

Steven Daly,[1] Frédéric Rosu,[2] Valérie Gabelica[1]*

[1]Université de Bordeaux, Inserm & CNRS, Laboratoire Acides Nucléiques: Régulations Naturelle et Artificielle (ARNA, U1212, UMR5320), IECB, 2 rue Robert Escarpit, 33607 Pessac, France.

[2] Université de Bordeaux, CNRS & Inserm, Institut Européen de Chimie et Biologie (IECB, UMS3033, US001), 2 rue Robert Escarpit, 33607 Pessac, France.

*Correspondence to: v.gabelica@iecb.u-bordeaux.fr.

**Materials and Methods**

Sample preparation

All DNA strands with D-sugars were purchased from Eurogentec (Liège, Belgium) with RP-cartridge purification and used as received. The L-dTGGGGT strand was purchased from Biomers Gmbh (Ulm, Germany) with HPLC purification and used as received. Nuclease-free water was from Ambion (Fischer Scientific, Illkirch, France). Ammonium acetate (NH$_4$OAc, BioUltra ~ 5M, Fluka), trimethylammonium acetate (TMAA, Ultra for UPLC ~ 1M, Fluka), and potassium (KCl, > 99.999%, Sigma) were purchased from Sigma-Aldrich (Saint-Quentin Fallavier, France).

The single strand dTGGGGT was dissolved in nuclease free H$_2$O to a concentration of 800 µM. This was diluted to 200 µM in 150 mM aqueous NH$_4$OAc and left overnight at 4 °C to yield a 50 µM quadruplex stock solution. This forms the [(dTGGGGT)$_4$•(NH$_4^+$)$_3$] tetramolecular G-quadruplex, abbreviated **TG4T**. For use in the ESI source, the stock solution was diluted to 5 µM G-quadruplex in 100 mM aqueous NH$_4$OAc.

The single strand d(T(GGT)$_4$TG(TGG)$_3$TGTT) was dissolved in nuclease free H$_2$O to a concentration of 200 µM. This solution was diluted to 50 µM in 150 mM aqueous NH$_4$OAc and left overnight at 4 °C to yield a stock solution. This forms the [dT(GGT)$_4$TG(TGG)$_3$TGTT•(NH$_4^+$)$_3$] intramolecular G-quadruplex, abbreviated **ZG4**. For use in the ESI source, the stock solution was diluted to 5 µM in 150 mM NH$_4$OAc.

The single strand dG$_{11}$ was dissolved in nuclease free H$_2$O to a concentration of 200 µM. AgNO$_3$ was prepared in H$_2$O at 1 mM. A stock solution with 50 µM dG$_{11}$ and 650 µM AgNO$_3$ was prepared in 10 mM NH$_4$OAc, giving a small excess of Ag$^+$ compared to the number of guanine base pairs. The stock solution was left at room temperature overnight to allow the formation of the duplex ([(dG$_{11}$)$_2$•(Ag$^+$)$_{11}$], abbreviated **GAgG**. For use in the ESI source, the stock solution was diluted to 5 µM duplex in 10 mM NH$_4$OAc.

The single strand d(GGGTTAGGGTTAGGGTTTGGG) was dissolved in nuclease free H$_2$O to a concentration of 200 µM. A stock solution of 50 µM strand in 1 mM KCl 100 mM trimethylammonium acetate (TMAA) was annealed at 80°C for 5 minutes and left overnight at room temperature to fold into a quadruplex. This forms the [d(GGGTTA)$_2$GGGTTTGGG•(K$^+$)$_2$] intramolecular G-quadruplex, abbreviated **5YEY**. For use in the ESI source, the quadruplex was diluted to 5 µM in 1 mM KCl and 100 mM TMAA.

Gas phase experimental setup

The experimental setup used for all gas phase experiments is shown in Fig. S4. The light source is a Spectra Physics PRO-230-30 Nd:YAG laser pumping a GWU Premiscan OPO running at 30 Hz, producing UV photons between 295 and 230 nm for our experiments. The laser is guided through a telescope, using an achromatic doublet (f = 200 mm) and a fused silica lens (f = -100 mm). The lenses are separated by 130 mm to obtain a collinear beam. Neutral density filters (0 to 0.5 optical density, combined range between 0 and 1.0) are fitted in two separate wheel mounts before and after the telescope, allowing to control the pulse energy. The beam then passes through an electro-mechanical shutter (SRS 470), which controls the number of pulses used to irradiate the ions. The shutter is synchronized to the laser and mass spectrometer cycle to ensure irradiation with a single pulse during the MS/MS activation stage of the mass spectrometer.

Next, the laser is guided through cage-mounted polarization optics. Firstly, a thin fused silica blade is used to reflect a portion of the beam to a pyroelectric energy meter (Ophir PE9-C), which can be used to constantly monitor the pulse energy during a measurement. The laser then passes through an air-spaced Rochon prism (Kogakugiken, Japan), and the extraordinary ray is blocked using an iris which allows the ordinary ray to pass. The laser then passes through a broadband UV quarter waveplate (Kogakugiken, Japan), which is held within a motorized rotation mount. Finally, to calibrate the reflected beam energy versus the energy of the beam used to irradiate the ions, a second pyroelectric energy meter (Ophir PE10) can be moved into the beam to measure the pulse energy following the polarization optics and prior to entering the mass spectrometer. The calibration is done prior to every measurement, and a typical example is shown in Fig. S8. The beam passes through a fused silica window into the pass spectrometer, and through 1.7 mm holes in the ring electrode of the ion trap. The beam is reflected by two 45° angle mirrors, and exits the mass spectrometer via a second fused silica window to avoid back scatter of photons within the trap.

The polarization optics and the two irises are connected to the same cage system. This cage system is in turn connected to a high load pitch and yaw platform. This allows the entire polarization optical setup to be moved relative to the beam without changing the laser alignment. To align the laser and polarization optics, first the laser is aligned to the ion trap by monitoring electron detachment yields of $[(TGGGGT)_4 \bullet (NH_4^+)_3]^{5-}$. Next, the caged-mounted polarization optics are positioned so that the laser passes through the center of the two irises, to achieve close to normal incidence of the laser through the polarizer and quarter waveplate. This process is iterated until the laser is aligned to the ions, and the polarization optics are aligned to the laser.

The next step is to align the fast axes of the polarizer and the quarter waveplate. To do this, we first orient the linear polarizer to maximize the pulse energy in the ordinary ray. Next, we take advantage of a dependence of the pulse energy transmitted by the quarter waveplate on the rotation angle of the quarter waveplate fast axis at 260 nm, which shows a maximum when the fast axes are aligned, and a minimum when at 90°. The rotation angle of the quarter waveplate is changed from -90° to 90° in 15° steps, and the transmitted and reflected pulse energies measured for 30 seconds. The zero offset of the motorized rotation mount is then optimized to give a maximum in the ratio of transmitted to reflected pulse energy at 0°, Fig. S5.

The beam then passes into the mass spectrometer (AmaZon, Bruker Daltonics, Germany) through 1.7 mm holes in the ring electrode. The ions are produced by electrospray ionization of solutions as described above. The ions are guided into the ion trap, accumulated for between 80 and 90 ms and mass selected with an m/z window of 5 Da where they are stored for 40 ms. During the trapping time, the shutter is opened to admit a single laser pulse into the mass spectrometer. The pulse energy is chosen so that only loss of a single electron is observed in the mass spectrum, and so that the electron detachment yield is still linear with respect to pulse energy, see Fig. S3.

Polarimetry

The polarization state of the laser in the ion trap is measured by polarimetry, using a system of polarizers that duplicates the experimental configuration as closely as possible, Fig. S6. An additional prism was added just after the mechanical shutter to guide the laser through the polarimeter. The polarimeter is comprised of four optical elements. The first linear polarizer and quarter waveplate (which performance is being characterized) are the same as used for CDMS experiments. Next, a second achromatic broadband quarter waveplate (B. Halle, Germany) is placed in a second motorized rotation mount. Finally, an alpha-BBO Glan-laser polarizer is placed, with a rotation angle of 0° that remains fixed throughout all polarimetry measurements. A pyroelectric energy meter (Ophir PE9-C) is placed after the polarimetry setup to measure the pulse energy. Between the two quarter waveplates, a fused silica window identical to that in the vacuum manifold of the mass spectrometer is used. The separation of the first polarizer, quarter waveplate and fused silica window are as close to real experimental conditions as possible.

The fast axis of the first air-spaced Rochon prism is used to define 0° for the other waveplates. First, the fast axis of the alpha-BBO Glan-laser polarizer was aligned. This was performed by finding the rotation angle of the alpha-BBO Glan-laser polarizer which maximized the pulse energy, without the quarter waveplates present. The angle of the linear polarizers is then fixed throughout. Next, we aligned the quarter waveplates individually. The first quarter waveplate was added to the system. The fast axis angle was then rotated from -90° to 90° and the pulse energy monitored, and the maximum denotes a fast axis angle of 0°. The second quarter waveplate was then added, and the same procedure was used. This ensured that all the optical elements where aligned to the fast axis of the linear polarizer.

In order to measure the polarization state of the laser, the rotation angle $\phi$ of the first quarter waveplate is fixed to a specific value. The rotation angle $\theta$ of the second quarter waveplate is changed between -90° and 90° in 15° steps, and the intensity of the transmitted light averaged for 30 seconds. The intensity of the laser pulse after traversing the polarimeter is given by *(32)*.

$$I(\theta) = \frac{1}{2}(A + B \sin 2\theta + C \cos 4\theta + D \sin 4\theta) \quad (1)$$

Where the coefficients A, B, C and D are related to the Stoke's parameters $S_0$, $S_1$, $S_2$ and $S_3$ by:

$$A = S_0 + \frac{S_1}{2}, B = S_3, C = \frac{S_1}{2}, D = \frac{S_2}{2} \quad (2)$$

The polarization state of the laser was measured at 295 nm, 260 nm and 240 nm. These wavelengths were chosen because they are diagnostic for DNA G-quadruplex topologies. The results are shown in Fig. S7 for $\phi = \pm 45°$ of the first quarter waveplate. Each curve has a dominant $\sin 2\theta$ component, which indicates highly pure circularly polarized radiation. The $S_3$ parameters extracted from the fit of this data are shown in Table S1. The $S_3$ parameter is greater than 0.95 for all wavelengths for $\phi = \pm 45°$, indicating that the polarization is at least 95% circularly polarized. The sign of the $S_3$ parameter identifies the helicity of the circular polarization. An angle of 45° gives left-handed circular polarized light, and -45° gives right-handed circular polarized light.

Finally, the stability and accuracy of the motorized rotation mount was tested. This is critical for circular dichroism measurements, where the quarter waveplate will be switched between $\phi = +45°$ and $\phi = -45°$ frequently. Polarimetry measurements were made by switching between 45° and -45° and repeating five times, Fig. S7. It can be seen that there is essentially no differences between the different repeated measurements. We can conclude that the motorized rotation mount is able to reproducibly reach the same rotation angle.

Circular dichroism mass spectrometry

Circular dichroism measurements in solution are performed by measuring the absorption using alternating pulses of left- and right-handed circularly polarized light. For gas phase measurements, we use action spectroscopy, and monitor the relative electron detachment efficiency, defined as:

$$Y_{ePD} = \frac{I_{ePD}}{I_{total}} / (\lambda * E) \quad (3)$$

where $I_{ePD}$ is the area under the peak in the mass spectrum due to electron detachment, $I_{total}$ is the total area under the mass spectrum, $\lambda$ is the wavelength and $E$ is the pulse energy.

First, we must find the optimal conditions to reproducibly measure the relative electron detachment efficiency. Two competing factors must be considered. One the one hand, it is desirable to change the polarization as often as possible, in order to minimize changes in the overlap of the ion cloud and laser, or changes in the pulse energy or beam profile. On the other hand, it is necessary to acquire mass spectra long enough to obtain an accurate measurement of the relative electron detachment efficiency. We found that a 90 second acquisition time was optimal to balance these two considerations.

The methodology used to perform circular dichroism measurements is as follows. The parent ion is generated and trapped as described in the experimental section. The wavelength and polarization state of the laser is fixed, and the trapped ions are irradiated with a single laser pulse. The mass spectrum is averaged for 90 seconds. As noted above, the ratio of transmitted to reflected pulse energy is sensitive to the rotation angle of the quarter waveplate. Therefore, the reflected pulse energy cannot be used alone for normalization. Thus, at each wavelength a calibration of transmitted and reflected pulse energy is performed prior to recording mass spectra. The rotation angle of the quarter waveplate is fixed to 45°, and the reflected and transmitted pulse energy averaged for a period of 30 seconds for a series of neutral density filters between 0 and 1 optical density in 0.1 steps. This process is then repeated with the quarter waveplate rotation angle fixed to 315°. Fig. S8 shows the reflected vs transmitted pulse energy for 45° and -45°, which have a linear correlation. The data is fitted with a linear function of the form $y = ax + b$, which is used to convert the reflected pulse energy recorded during a circular dichroism measurement to the transmitted pulse energy, which is then used for normalization.

Next, the quarter waveplate rotation angle is fixed to 45°, the trapped ions are irradiated and the mass spectrum averaged for 90 seconds. The rotation angle is then changed to -45° and the mass spectrum averaged for 90 seconds. This process is repeated 10 times, giving a total irradiation time of 15 minutes for both 45° and -45° (right and left CP respectively). This process is then repeated for each wavelength used.

In order to extract the circular dichroism value, the relative electron detachment efficiency is determined for each mass spectrum acquired above, giving 10 values for RCP and LCP using equation 3. The average value of the relative detachment efficiency for RCP and LCP is determined.

$$\overline{Y_{ePD}^{RCP}} = \frac{\sum_n Y_{ePD}^{RCP}}{n} \quad (4)$$

$$\overline{Y_{ePD}^{LCP}} = \frac{\sum_n Y_{ePD}^{LCP}}{n} \quad (5)$$

Circular dichroism is defined as the difference in absorption of left- and right-handed circular polarized photons. Here, we can make the assumption that the relative electron detachment efficiency is directly correlated to the absorption cross-section. Following the convention for circular dichroism measurements, we define the asymmetry factor $g_{ePD}$ as:

$$g_{ePD} = \frac{\Delta ePD}{ePD} = 2 * \left(\frac{\overline{Y_{ePD}^{LCP}} - \overline{Y_{ePD}^{LCP}}}{\overline{Y_{ePD}^{LCP}} + \overline{Y_{ePD}^{LCP}}}\right) \quad (6)$$

This gives a unitless value for the circular dichroism, which can be directly compared to solution phase measurements.

$$CD = g_{sol} = \frac{\Delta A}{\bar{A}} = \frac{\Delta \epsilon}{\bar{\epsilon}} \qquad (7)$$

Where $A$ is the absorbance and $\epsilon$ is it extinction coefficient.

In order to determine the reproducibility of circular dichroism mass spectrometry measurements, two different methodologies were used. Firstly, for each species the circular dichroism measurements are repeated 3 times for at least wavelength. The standard error in the repeated measurements is then determined and reported.

Solution Phase circular dichroism

All solution phase measurements were performed on a JASCO 1500 circular dichroism spectrometer. The samples for the circular dichroism measurements were prepared in an identical manner as those used in mass spectrometry, as described above. Spectra were recorded between 350 and 220 nm, and blank subtractions performed. For data analysis, the circular dichroism in mdeg was first converted to ΔAbs using the conversion factor:

$$CD(\Delta Abs) = \frac{CD(mdeg)}{32980} \qquad (8)$$

The asymmetry factor as then calculated as

$$g_{sol} = \frac{\Delta A}{A} \qquad (9)$$

Ion Mobility Mass Spectrometry

The experiments were carried out on an Agilent 6560 IMS-Q-TOF (Agilent Technologies, Santa Clara, CA), with its drift tube ion mobility cell operated in helium. The instrument tuning parameters are the same as the "compromised" tuning parameters described and discussed elsewhere (penultimate column of Tables 1 and 2 in (*33*)). The helium pressure in the drift tube was 3.89 ± 0.01 Torr, and the pressure in the trapping funnel is 3.63 ± 0.01 Torr. The pressure differential between the drift tube and the trapping funnel ensures only helium is present in the drift tube. Injection was in negative ion mode, using the standard electrospray source and a syringe pump at 4 µL/min. The acquisition software version was B.07.00 build 7.00.7008. The arrival time distributions were extracted from the entire isotopic distribution of each adduct, using IM-MS browser.

Step-field experiments (five drift tube voltages for each samples) were performed to determine the CCS. The arrival time distributions (ATDs) for each charge state of the complexes were fitted with one gaussian peak using OriginPro 2016, to determine the arrival time $t_A$ of the center of the peak. The arrival time $t_A$ is related to $\Delta V$ (voltage difference between the entrance and the exit of the drift tube region) by:

$$t_A = \frac{L^2}{K_0} \frac{T_0 p}{p_0 T} \cdot \left(\frac{1}{\Delta V}\right) + t_0 \qquad (10)$$

$t_0$ is the time spent outside the drift tube region and before detection. A graph of $t_A$ vs. $1/\Delta V$ provides $K_0$ from the slope and $t_0$ as the intercept. The drift tube length is $L = 78.1$ cm, the temperature is measured accurately by a thermocouple (here, $T = 297 \pm 1$ K), and the pressure is measured by a capacitance gauge (p = $3.89 \pm 0.01$ Torr). The CCS is then determined using Equation (11):

$$CCS = \frac{3ze}{16N_0} \cdot \sqrt{\frac{2\pi}{\mu k_B T}} \cdot \frac{1}{K_0} \qquad (11)$$

The reconstruction of the experimental CCS distributions from the arrival time distributions at the lowest voltage is then performed using Equation (12), where the factor $a$ is determined from the $t_A$ of the peak center at the lowest voltage and the $CCS$ calculated from the regression described above, from the peak centers.

$$CCS = a \cdot \frac{z}{\sqrt{\mu}} \times t_A \qquad (12)$$

Generation of gas-phase structures and calculation of theoretical collision cross sections

The gas-phase modeling was started from the published structure (PDB code: 1S45 for TG4T, 5YEY for the antiparallel G-quadruplex, 4U5M for the Z-G4 and ref (*20*) for GAgG11). Phosphate groups (arbitrarily chosen) where neutralized by protons to attain a total charge state of 5- or 6-. We optimized the structure at the semi-empirical level with PM7 (*34*) parameterization, using Gaussian 16 rev. B.01 (*35*). Then, atom Centered Density Matrix Propagation molecular dynamics (ADMP, 1000 fs, 296 K) at the semi-empirical level (PM7) was performed. The theoretical CCS values were calculated for a structure every 10 fs, using the trajectory model (Mobcal (*36*), original parameters for helium, N and O parameterized as C, P and K parameterized as Si). The GAgG11 structure is calculated as described in ref (*20*).

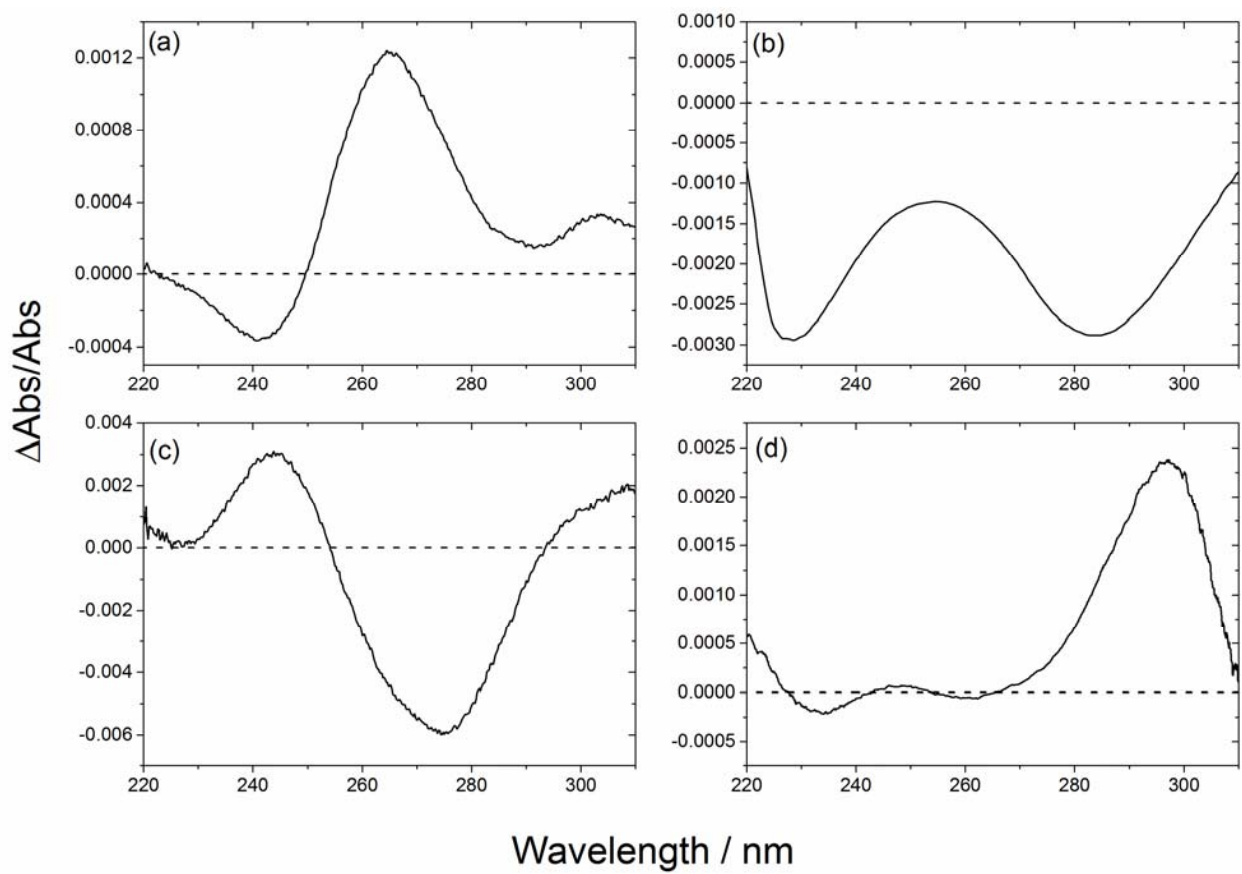

**Fig. S1. Solution phase electronic circular dichroism spectra (all D-sugars).**
(a) TG4T (5 μM tetramolecular G-quadruplex, 100 mM ammonium acetate), (b) GAgG (5 μM duplex, 60 μM AgNO$_3$, 20 mM ammonium acetate), (c) ZG4 (5 μM dT(GGT)$_4$TG(TGG)$_3$TGTT, 100 mM ammonium acetate) and (d) 5YEY (5 μM d(GGGTTA)$_2$GGGTTTGGG, 1 mM KCl, 100 mM trimethylammonium acetate).

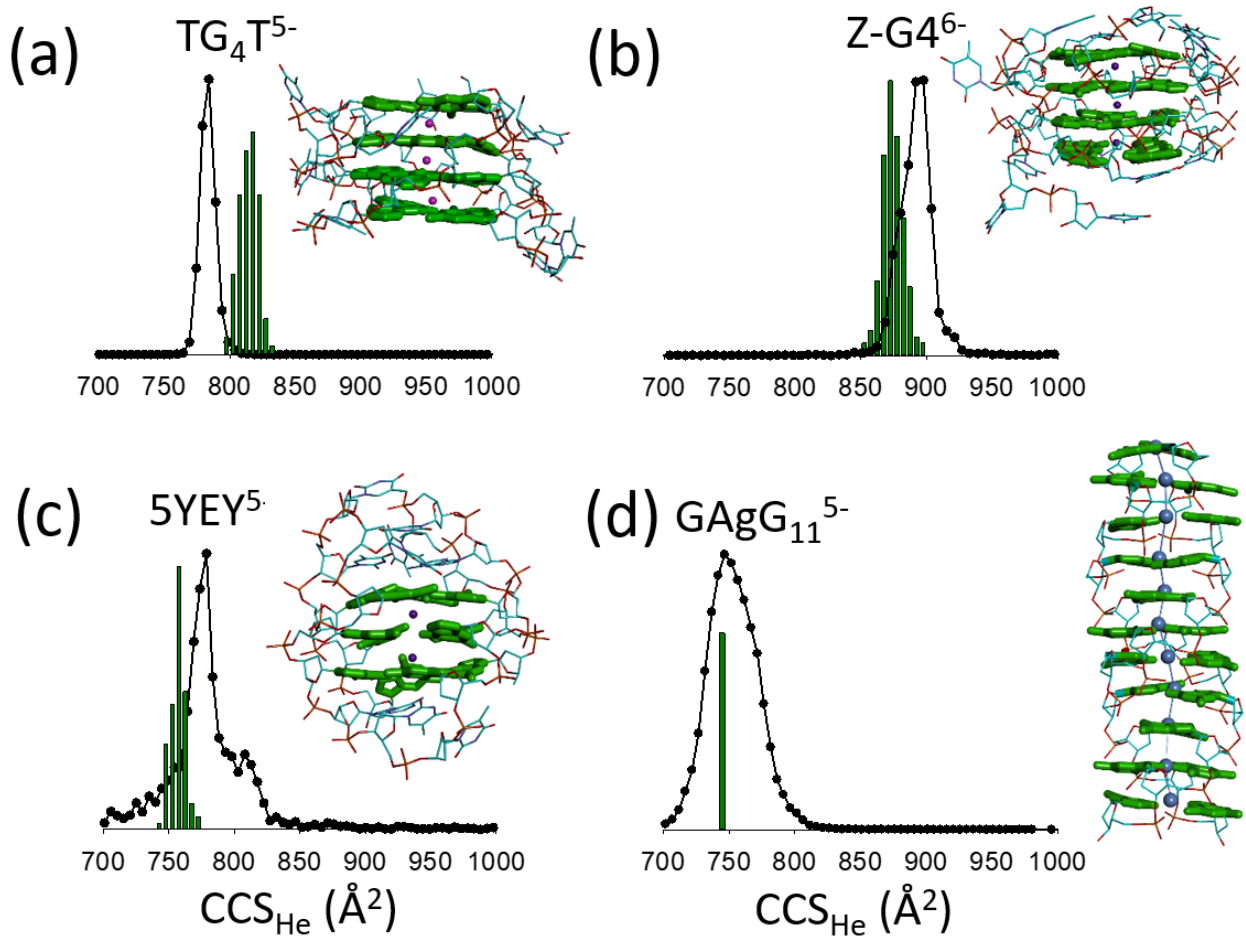

**Fig. S2. Gas-phase ion mobility spectrometry and calculated collision cross sections.**
Black dots: Experimental helium collision cross section distributions reconstructed for (a) $[(dTGGGGT)_4•(NH_4^+)_3]^{5-}$, (b) $[d(GGGTTA)_2GGGTTTGGG•(K^+)_2]^{5-}$, (c) $[dT(GGT)_4TG(TGG)_3TGTT•(NH_4^+)_3]^{6-}$ and (d) $[(dG_{11})_2•(Ag^+)_{11}]^{5-}$. Green bars: calculated helium collision cross sections for a collection of gas-phase structures generated by PM7 optimization followed by PM7 BOMD, except for GAgG (panel d) for which the structure was generated as described in (*20*), and PM7 BOMD was not feasible due to inappropriate parameterization of silver.

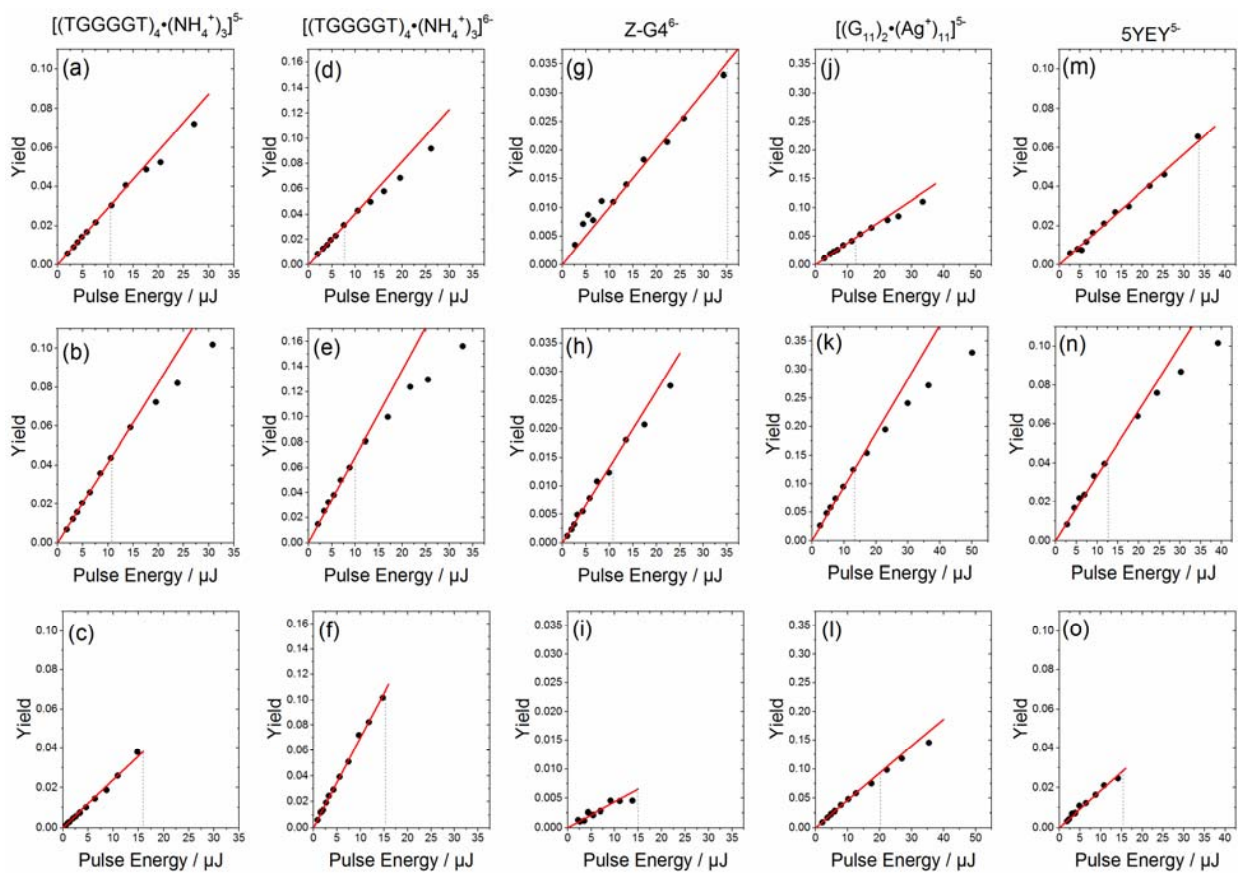

**Fig. S3. Linearity of the electron photodetachment process.**

Electron detachment yield ($I_{ePD}/I_{total}$) as a function of the laser pulse energy of TG4T$^{5-}$ ions at (a) 240 nm, (b) 260 nm and (c) 280 nm TG4T$^{6-}$ ions at (d) 240 nm, (e) 260 nm and (f) 280 nm. ZG4$^{6-}$ ions at (g) 240 nm, (h) 260 nm and (i) 280 nm. [(G$_{11}$)$_2$•(Ag$^+$)$_{11}$]$^{5-}$ ions at (j) 240 nm, (k) 260 nm and (l) 280 nm. 5YEY$^{5-}$ ions at (m) 240 nm, (n) 260 nm and (o) 280 nm. The red line represents a linear fit to the data passing through the origin.

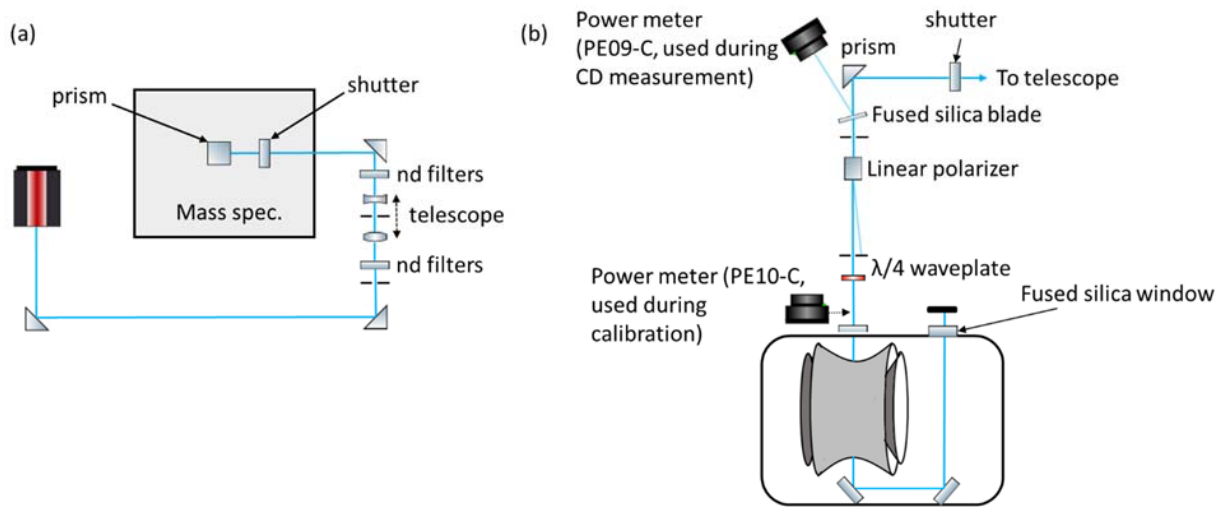

**Fig. S4. Experimental setup used for circular dichroism mass spectrometry.**
(a) Top view showing the laser, the neutral density filters used to control pulse energy and the telescope used to collimate the beam. (b) Side view of the polarization optics. The beam passes through a shutter, is reflected through 90° by a prism and through a fused silica blade. This reflects a small portion of the beam to a power meter (PE09-C) which is used to record the pulse energy during mass spectrum acquisitions. The beam then passes through the polarization optics. A second pulse energy meter (PE10-C) can be inserted into the beam just before the mass spectrometer, and is used to convert reflected pulse energies to transmitted pulse energy for normalization. The beam then passes through a fused silica window, through a Paul trap, is reflected by two 45° mirrors and exits the mass spectrometer through a second fused silica window, where it is blocked.

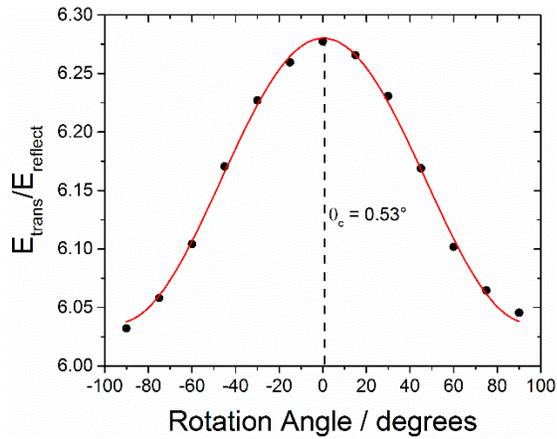

**Fig. S5. Verification of the quarter wave plate.**

Ratio of transmitted to reflected pulse energy as a function of the fast axis angle of the quarter waveplate (black squares) at 260 nm. The red line is a fit to a cosine function $y = y_0 + A\cos\left[\frac{(x-x_c)\pi}{p}\right]$. The value $x_c$ gives the offset between the fast axis of the quarter waveplate and the linear polarizer.

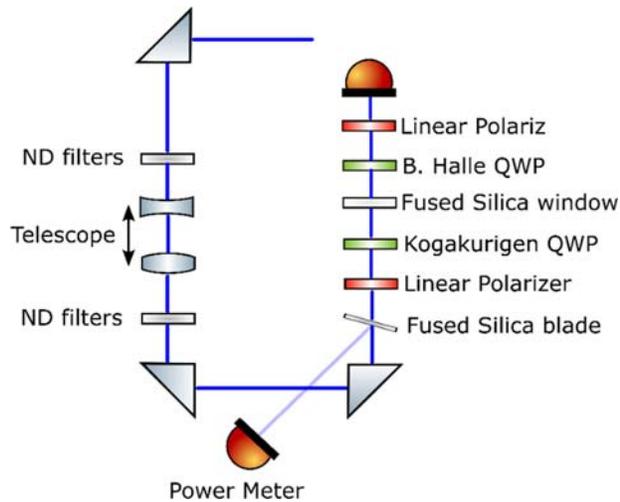

**Fig. S6. Polarimetry setup for confirming the polarization state of laser pulses.**

An additional prism is added after the shutter in Fig. S1. The fused silica blade, linear polarizer and first quarter waveplate are in an identical configuration as used in CDMS experiments. The fused silica window is identical to the one in the vacuum manifold to allow light to couple to the ion trap. The second quarter waveplate and linear polarizer are used to measure the polarization after the fused silica window.

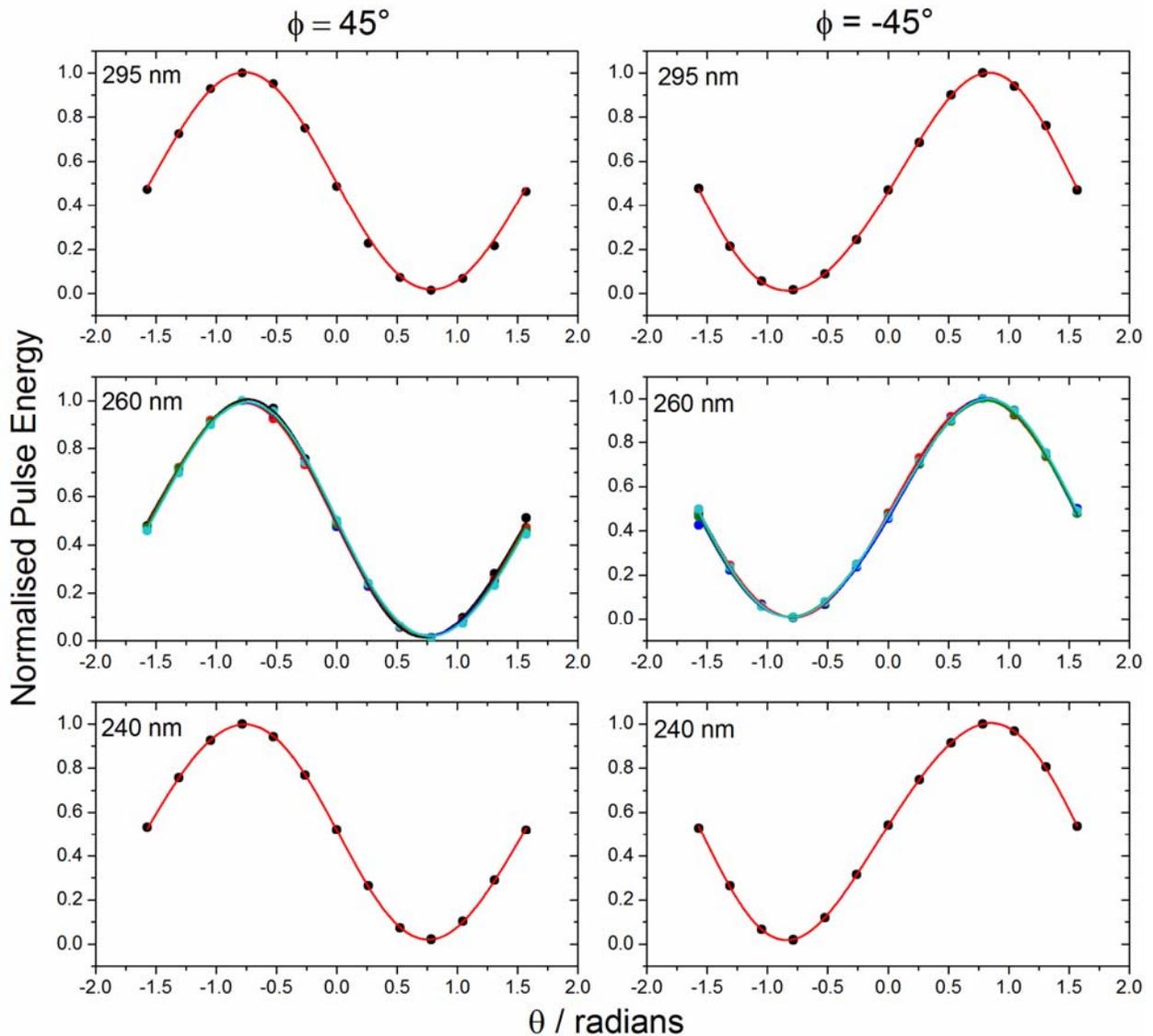

**Fig. S7. Polarimetry results.**

Normalized pulse energy (circles) as a function of the second quarter waveplate angle θ for $\phi = \pm 45°$ of the first quarter waveplate at 295 nm (top), 260 nm (middle) and 240 nm (bottom). The red curve shows the fit to the function in equation 1. For 260 nm, 5 measurements were made to show the reproducibility of the polarization and that there was no change when repeatedly cycling form 45° to -45° and back.

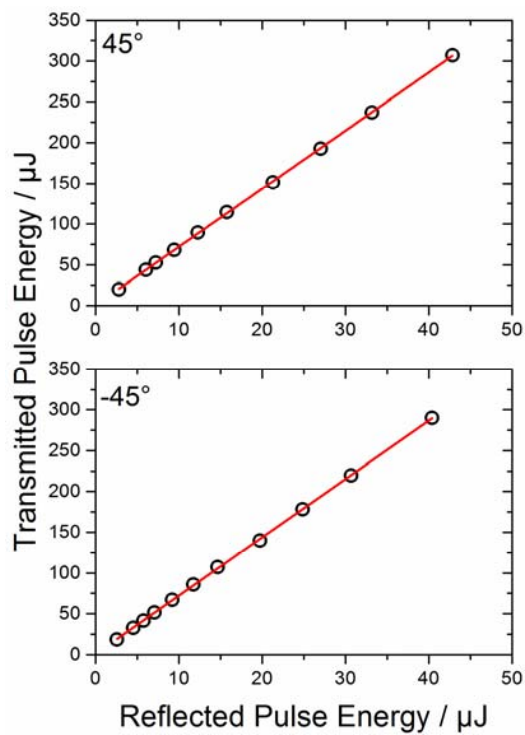

**Fig. S8. Calibration of the transmitted vs. reflected laser pulse energy.**

Transmitted versus reflected pulse energy at 260 nm for 45° (top) and -45° (bottom) rotation angle of the quarter waveplate, using neutral density filters to give optical densities between 0 and 1.0 (circles). The red line is a linear fit ($y = ax + b$) of the data, which is used to convert the reflected pulse energy averaged during measurement of a mass spectrum to the transmitted pulse energy that can be used for normalization in equation 3.

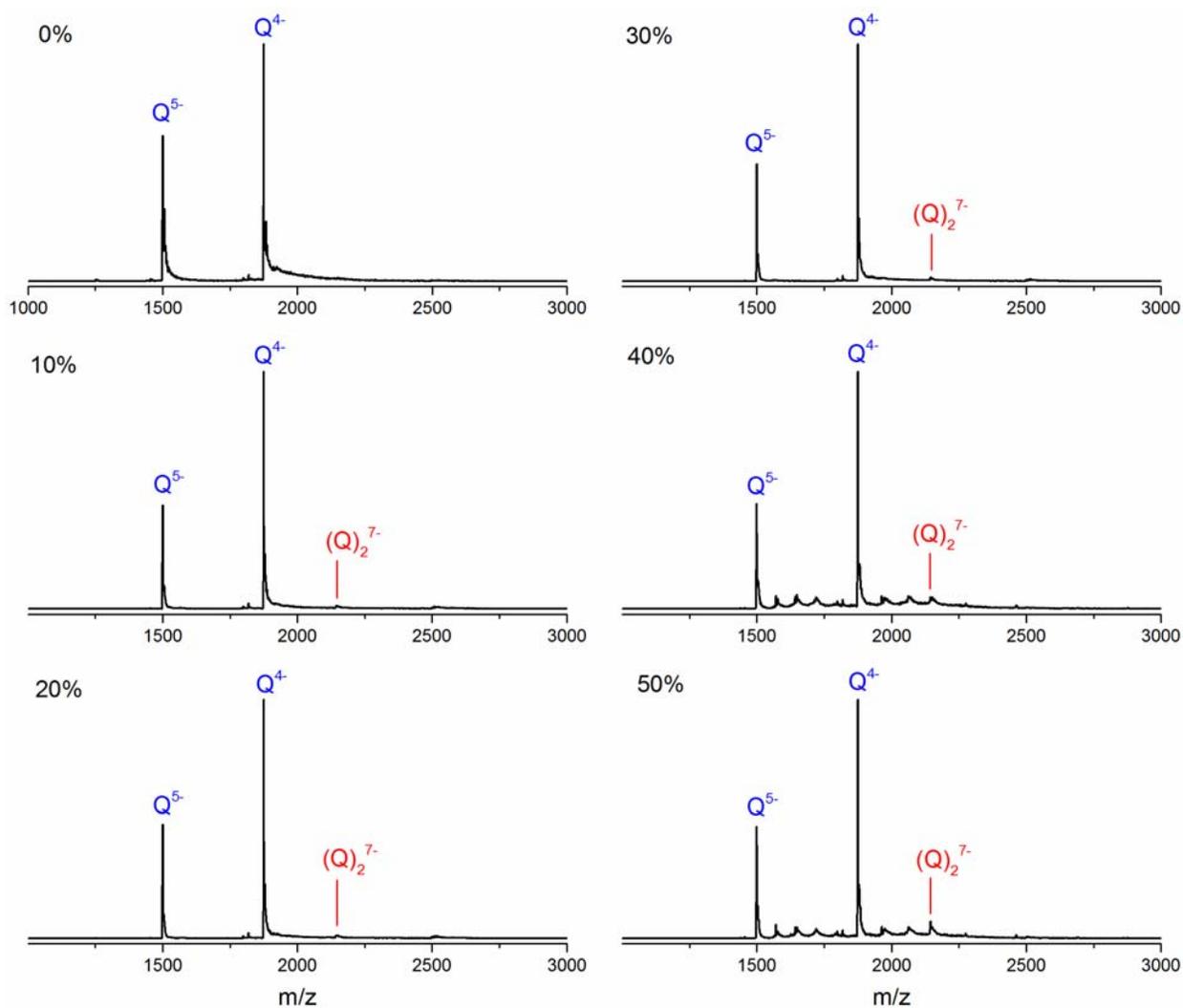

**Fig. S9. Addition of up to 50% isopropanol does not cause aggregation of TG4T.**
Mass spectra of TGGGGT in 100mM ammonium acetate at different percentages v/v of isopropanol. Q denotes the intact quadruplex structures [(TGGGGT)$_4$•(NH$_4^+$)$_3$], and S denotes the single strand. There is no significant change in the mass spectra as a function of the isopropanol concentration. Above 40%, some adducts appear, but there is no evidence for the formation of any higher order oligomers, or for significant disruption of the quadruplexes.

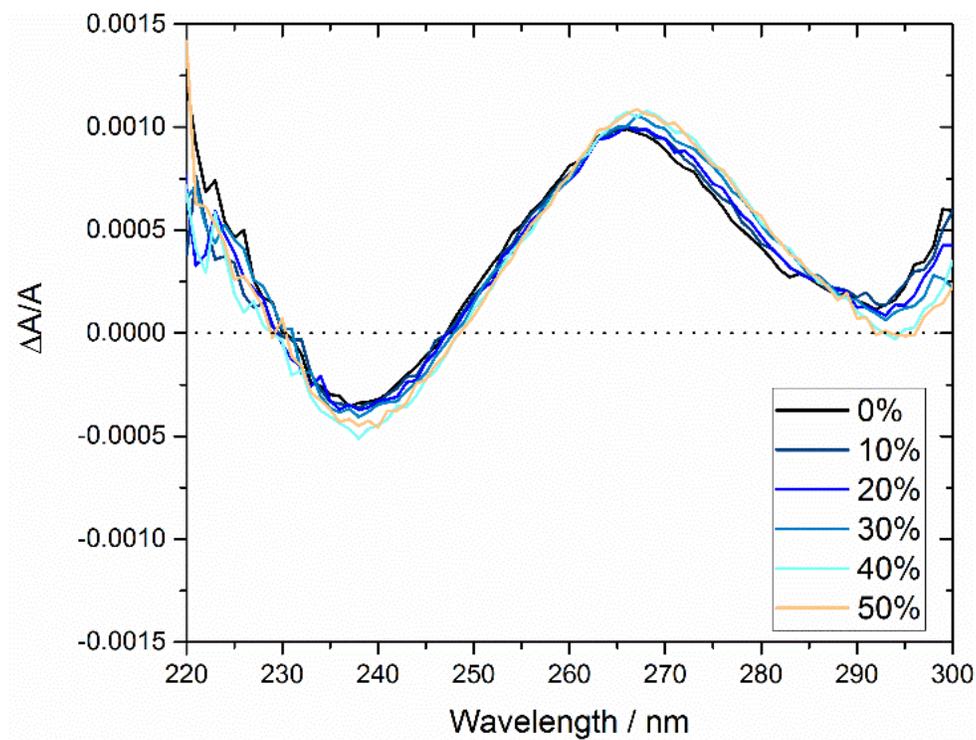

**Fig. S10. In solution, the solvent polarity does not affect the CD spectrum of TG4T.**
Circular dichroism spectra for 5μM $D$-[(TGGGGT)$_4$•(NH$_4^+$)$_3$] in 100 mM ammonium acetate for different volume percentages of isopropanol.

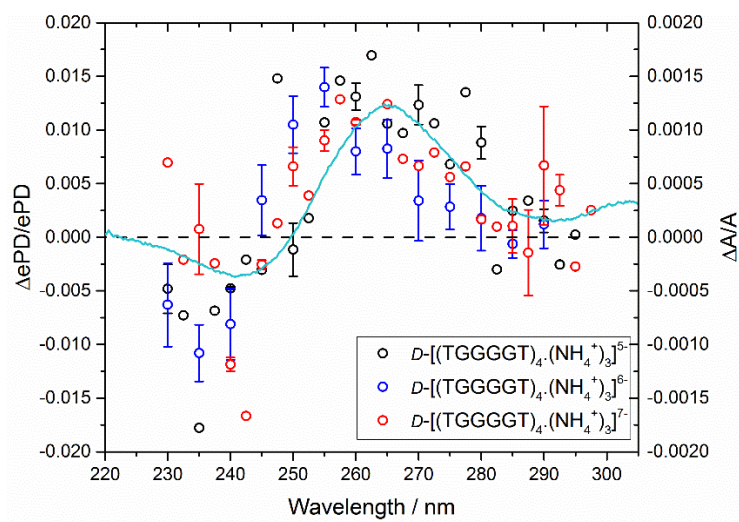

**Fig. S11. The charge state does not affect the magnitude of the gas-phase CD effect.**
Gas phase circular dichroism (circles) for D-[(TGGGGT)$_4$•(NH$_4^+$)$_3$] in the 5- (black), 6- (red) and 7- (blue) charge states. Errors bars represent the standard error in 3 repeated measurements. The cyan curve shows the solution phase circular dichroism of D-[(TGGGGT)$_4$•(NH$_4^+$)$_3$] in 100 mM ammonium acetate.

**Table S1. S$_3$ parameters for the data shown in Fig. S6.**

The values in parentheses represent the error in the fitting procedure for 240 nm and 295 nm. For 260 nm, the values are the standard error in the 5 repeated measurements.

|        | 45° (LCP)  | -45° (RCP) |
|--------|------------|------------|
| **240 nm** | -0.963(6)  | 0.962(5)   |
| **260 nm** | -0.960(2)  | 0.980(1)   |
| **295 nm** | -0.965(22) | 0.969(9)   |

**Supplementary references**

20. S. M. Swasey, F. Rosu, S. M. Copp, V. Gabelica, E. G. Gwinn, *J. Phys. Chem. Lett.* **9**, 6605-6610 (2018).
32. B. Schaefer, E. Collett, R. Smyth, D. Barrett, B. Fraher, *Am. J. Phys.* **75**, 163-168 (2007).
33. V. Gabelica, S. Livet, F. Rosu, *J. Am. Soc. Mass Spectrom.* **29**, 2189-2198 (2018).
34. J. J. Stewart, *J. Mol. Model.* **19**, 1-32 (2013).
35. M. J. Frisch *et al.*, *Gaussian 16 Rev. B.01*. (Wallingford, CT, 2016).
36. M. F. Mesleh, J. M. Hunter, A. A. Shvartsburg, G. C. Schatz, M. F. Jarrold, *J. Phys. Chem.* **100**, 16082-16086 (1996).